%% file: TOP-17-016_temp.tex
\pdfoutput=1

\documentclass[11pt,twoside,a4paper,cmspaper,final,collab]{cms-tdr}
\def\svnVersion{483218P}\def\cmsCernNoTag{CERN-EP-2018-206}\def\cmsCernDate{\today}\def\cmsMessage{Published in Physical Review Letters as \href{http://dx.doi.org/10.1103/PhysRevLett.121.221802}{\doi{10.1103/PhysRevLett.121.221802.}}}

\begin{document}\cmsNoteHeader{TOP-17-016}

\hyphenation{had-ron-i-za-tion}
\hyphenation{cal-or-i-me-ter}
\hyphenation{de-vices}
\RCS$HeadURL: svn+ssh://svn.cern.ch/reps/tdr2/papers/TOP-17-016/trunk/TOP-17-016.tex $
\RCS$Id: TOP-17-016.tex 483110 2018-12-02 11:57:19Z setesami $

\newlength\cmsFigWidth
\ifthenelse{\boolean{cms@external}}{\setlength\cmsFigWidth{0.85\columnwidth}}{\setlength\cmsFigWidth{0.4\textwidth}}
\ifthenelse{\boolean{cms@external}}{\providecommand{\cmsLeft}{top\xspace}}{\providecommand{\cmsLeft}{left\xspace}}
\ifthenelse{\boolean{cms@external}}{\providecommand{\cmsRight}{bottom\xspace}}{\providecommand{\cmsRight}{right\xspace}}

\newcommand{\wjets}{\ensuremath{\PW}+jets\xspace}
\newcommand{\wajets}{\ensuremath{\PW\gamma}+jets\xspace}

\newcommand{\zjets}{\ensuremath{\cPZ}+jets\xspace}
\newcommand{\zajets}{\ensuremath{\cPZ\gamma}+jets\xspace}

\newcommand{\ttbara}{\ensuremath{\ttbar}+\ensuremath{\gamma}\xspace}

\cmsNoteHeader{TOP-17-016}

\title{Evidence for the associated production of a single top quark and a photon in proton-proton collisions at  \texorpdfstring{$\sqrt{s}=13\TeV$ }{sqrt(s) = 13 TeV}}

\date{\today}

\abstract{
The first evidence of events consistent with the production of a single top quark in association with a photon is reported.
The analysis is based on proton-proton
collisions at $\sqrt{s}=13\TeV$  and recorded by the CMS experiment in 2016,
corresponding to an integrated luminosity of 35.9\fbinv.
Events are selected by requiring the presence of a muon (\Pgm), a
photon ($\gamma$), an imbalance in transverse momentum from an undetected neutrino ($\nu$),
and at least two jets (j) of which exactly one is identified as
associated with the hadronization of a \PQb quark. A multivariate discriminant
based on topological and kinematic event properties is employed to
separate signal from
background processes. An excess above the
background-only hypothesis is observed, with a significance
of 4.4 standard deviations. A fiducial cross section is measured for
isolated photons with transverse momentum greater than 25\GeV in the central region of the detector.
The measured product of the cross section and branching fraction is $\sigma(\Pp\Pp\to\PQt\gamma\mathrm{j})\mathcal{B}(\PQt\to\Pgm\nu\PQb) = 115\pm 17\stat\pm 30\syst\unit{fb}$,  which is consistent with the standard model prediction.
}

\hypersetup{%
pdfauthor={CMS Collaboration},%
pdftitle={Evidence for the associated production of a single top quark and a photon in proton-proton collisions at  sqrt(s) = 13 TeV},%
pdfsubject={CMS},%
pdfkeywords={CMS, top quark, photon}}

\maketitle

The study of top quark production in association
with a photon is an important test of the standard model (SM)
description of top quark interactions with gauge bosons, and plays an
important role in the search for physics beyond the SM. The cross section for single top quark production in association
with a photon is sensitive to the top quark charge, and to its
electric and magnetic dipole moments~\cite{Fael:2013ira,Etesami:2016rwu}.

The cross section for the production of a top quark-antiquark pair
and an associated photon (\ttbara), has been measured
by the CDF Collaboration at the Fermilab Tevatron in proton-antiproton collisions at $\sqrt{s}=1.96\TeV$~\cite{Aaltonen:2011sp},
and by the ATLAS and CMS Collaborations in proton-proton (\Pp\Pp) collisions at the  CERN LHC at 8\TeV~\cite{Aaboud:2017era,Sirunyan:2017iyh}.
 All measurements are consistent with SM predictions.

\begin{figure*}[htb]
\centering
\includegraphics[height=0.22\textwidth]{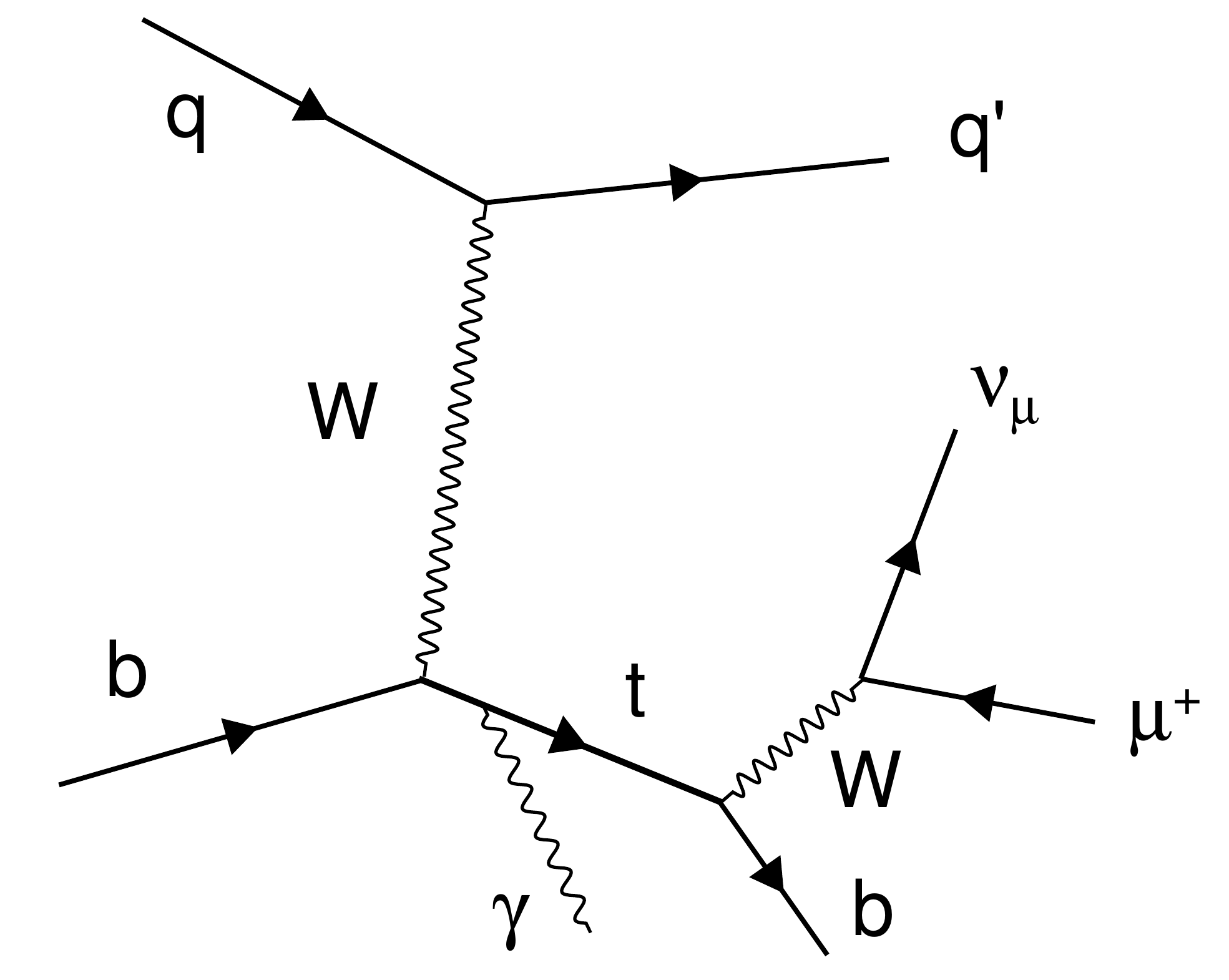}
\includegraphics[height=0.22\textwidth]{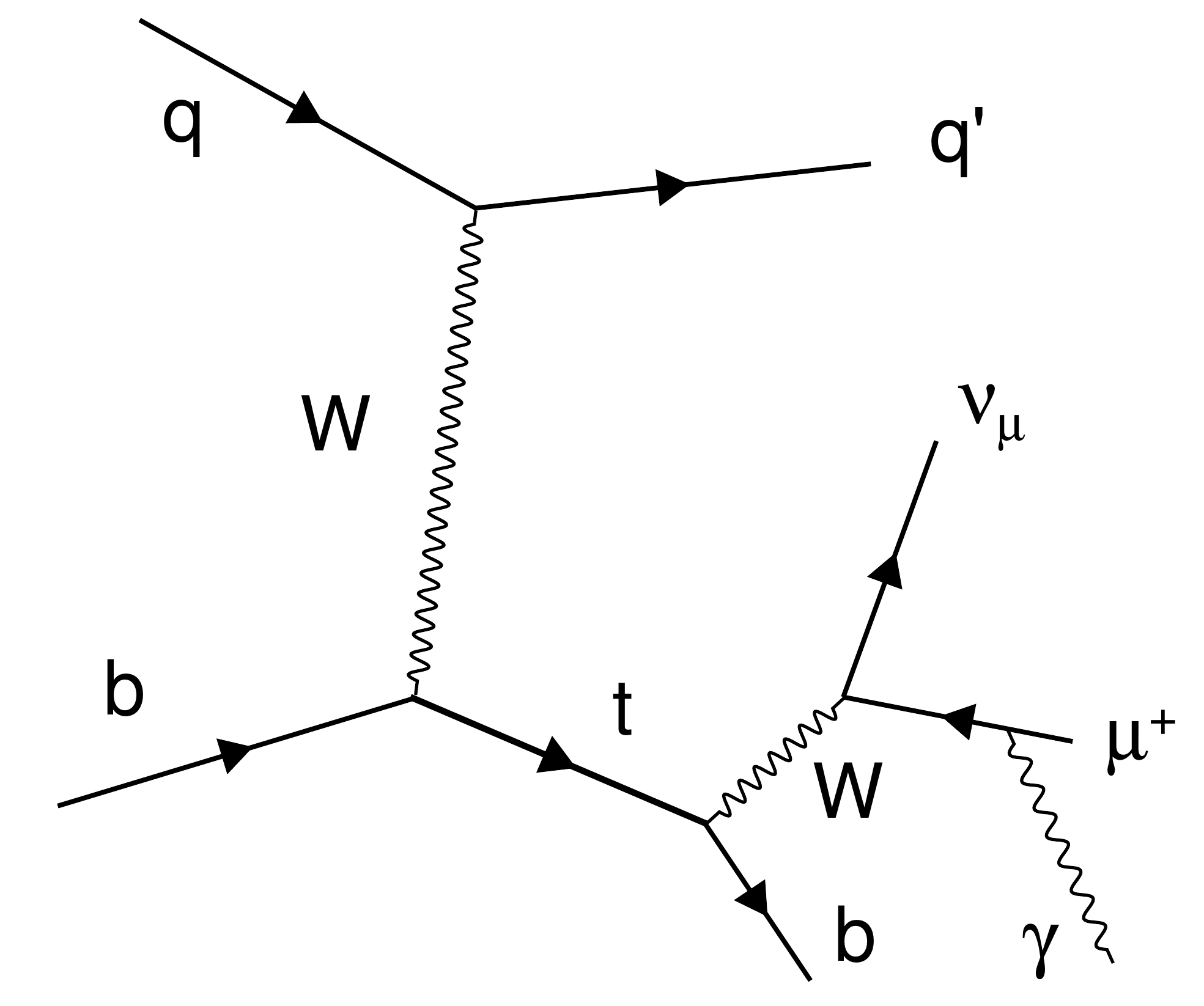}
\includegraphics[height=0.22\textwidth]{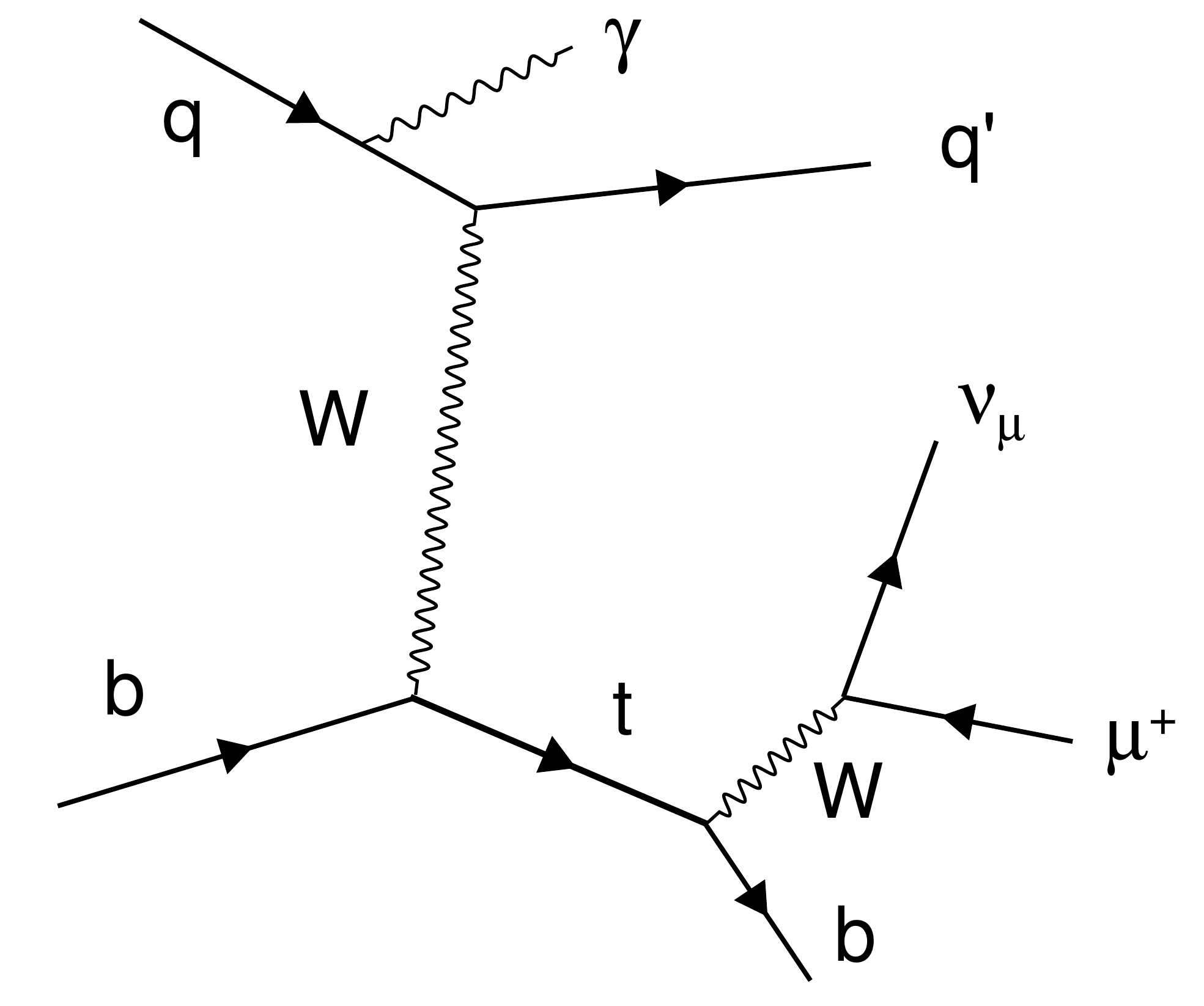}\\
\caption{ Representative $t$-channel Feynman diagrams for single top quark production
  in association with a photon, including the leptonic decay of the \PW\ boson produced  in the
top quark decay. }
\label{feynman}
\end{figure*}

The SM predicts  three different mechanisms for the production of a single top quark in association with a photon at the LHC:
$t$-channel, $s$-channel, and $\PQt\PW$ production, where the largest contribution comes from the
$t$ channel.
This Letter presents first evidence for single top quark
production in association
with a photon in the $t$ channel. The data were
collected in 2016 by the CMS experiment at
$\sqrt{s}=13\TeV$ and correspond to
an integrated luminosity of 35.9\fbinv.
One of the distinctive signatures of  the signal is the presence of a forward light-flavor energetic
jet in the final state that arises from the space-like nature of the \PW\ boson propagator.
The search targets events with a top quark, a photon, and at least one additional
light-flavor  jet, ($\PQt\gamma$j) where
the top quark decays into a \PW\ boson and a \PQb\ quark, followed by the \PW\ boson
decay into a muon and a neutrino.  Figure~\ref{feynman} shows representative Feynman diagrams for the
process $\PQt\gamma$j including the
 leptonic decay of the \PW\ boson produced in the top quark decay.

This analysis focuses on the muon decay channel since it enables a good signal
selection efficiency with low background contamination.
The final state includes possible contributions from $\PW\to\Pgt\nu_{\Pgt}$
where the \Pgt\ lepton decays to $\Pgm\nu_{\Pgm}\nu_{\Pgt}$.
The kinematic and topological properties of the signal are exploited through a multivariate technique to
 discriminate against the background processes.
The background can be divided into two types: events which contain a genuine photon, such as
\ttbara, \wajets, and \zajets and events in which the photon
 candidate arises from a misidentified jet, such as \ttbar, \wjets,
 and \zjets processes.

The central feature of the CMS apparatus is a superconducting solenoid
of 6\unit{m} internal diameter, providing a magnetic field of
3.8\unit{T}. A silicon
 pixel and strip tracker, a lead tungstate crystal electromagnetic
 calorimeter (ECAL), and a brass and scintillator hadron calorimeter, each composed
 of a barrel and two endcap sections, are contained within the superconducting solenoid volume.
 Extensive forward
 calorimetry complements the coverage provided by the barrel and endcap detectors.
Muons are detected in gas-ionization detectors embedded in the
 steel flux-return yoke outside the solenoid.
Online event selection is accomplished
via the two-tiered CMS trigger system~\cite{trigger}.
A more detailed description of the CMS detector, together with a
definition of the coordinate
 system and kinematic variables used in this analysis, can be found in
 Ref.~\cite{Chatrchyan:2008zzk}.

 Simulated samples for the  $\PQt\gamma$j signal are generated at next-to-leading order (NLO)
 using the \MGvATNLO v 2.2.2 event generator~\cite{Alwall:2014hca},
 with a minimum transverse momentum requirement of  $\pt>10\GeV$
 and a pseudorapidity requirement of $\abs{\eta}<2.6$
for the associated photon.  The angular separation between the photon and all other particles is
required to be $\DR>0.05$, where $\sqrt{\smash[b]{(\Delta\eta)^2+(\Delta\phi)^2}}$ and $\phi$ is the azimuthal angle in radians.
The NNPDF3.0~\cite{Ball:2014uwa} parton distribution functions (PDFs)
are used and the top quark mass is set to 172.5\GeV.
After these requirements, an inclusive cross
  section of $2.95\pm0.13\pm0.03\unit{pb}$ is obtained,
  where the first uncertainty is associated with the
  renormalization and factorization scales
  and the second uncertainty with the PDFs.
  These uncertainties are derived using the \textsc{SysCalc}
  program~\cite{Kalogeropoulos:2018cke}.

Samples of simulated events for the production of \ttbara,
\wjets, \wajets, Drell--Yan events, \zajets, and diboson+$\gamma$
are generated at NLO using \MGvATNLO. Single top quark events
in $t$-, $s$-, and $\PQt\PW$-channels, as well as \ttbar events, are generated
with the NLO \POWHEG v2 event generator
\cite{Nason:2004rx,Frixione:2007vw,Alioli:2010xd,Campbell:2014kua}.
The overlaps between the $\PQt\gamma$j signal sample and the tj inclusive
sample, and between V$\gamma$+jets and V+jets samples
are removed, where V is a \PW\ or \cPZ\ boson.

Showering and hadronization for all of the simulated samples are implemented
with \PYTHIA v8.212~\cite{Sjostrand:2006za, Sjostrand:2007gs} with the
underlying event tune CUETP8M1~\cite{Khachatryan:2015pea}. To match
the generated events with \PYTHIA v8.212 parton shower, the
FxFx~\cite{Alwall:2007fs} and the MLM~\cite{Frederix:2012ps}
prescriptions are used.

The CMS detector response for all simulated samples is modeled using
\GEANTfour v9.4~\cite{geant}. The simulated samples include the presence of additional \Pp\Pp\ interactions in the same or neighboring bunch crossings (pileup).
 The distribution of pileup events in simulation is weighted to match that
observed in data.

The particle-flow (PF) algorithm~\cite{Sirunyan:2017ulk} is used for
the identification and
reconstruction of individual particles in an event using a combination of information
from all subdetectors.  It identifies each reconstructed particle as a muon, an electron,
a photon, a neutral or a charged  hadron.
Data samples are selected based on a single-muon trigger, requiring a
muon with $\pt >24\GeV$ and $\abs{\eta}<2.4$. Events are required to have
a well-reconstructed primary vertex which is identified as the one
with the largest value of summed particle $\pt^2$~\cite{Chatrchyan:2014fea}. The simulated events are weighted in order to reproduce the reconstruction  and trigger efficiencies
in data.

Jets are reconstructed from the PF candidates using the anti-\kt algorithm~\cite{Cacciari:2008gp} with a distance
parameter R = 0.4, implemented in the \FASTJET package~\cite{Cacciari:2011ma}. Corrections are applied to
the jet energy to subtract the contribution
from pileup interactions. Jets are
tagged as associated to the hadronization of \PQb\ quarks (\PQb\ jets) using
secondary vertex algorithm, which combines the secondary vertex and track-based
lifetime information~\cite{Sirunyan:2017ezt}. The working point chosen
for the algorithm, corresponds to a \PQb\ tagging efficiency
of 70\%,  and  misidentification rates of 1  and 15\% for light quark
jets and \PQc jets, respectively~\cite{Sirunyan:2017ezt}.
Events are required to have at least two jets with $\pt>40\GeV$,
one of which must be $\PQb$-tagged with $\abs{\eta}<2.5$ and another must lie
within the range $\abs{\eta}<4.7$.

Events are required to contain exactly one isolated
muon with $\pt>26\GeV$ and $\abs{\eta}<2.4$.
The isolation is calculated from the reconstructed charged and neutral PF
candidates and is corrected for pileup effects~\cite{Sirunyan:2018fpa}.
In order to reduce the contribution of background processes with multiple leptons in the final state, events
containing additional muon candidates satisfying loose selection criteria or containing electron candidates
are rejected~\cite{Sirunyan:2018fpa}.

Photon candidates are built from clusters of high-energy deposits in the ECAL.
Photon identification depends on isolation and shower shape variables which reflect
the energy dispersion in $\eta$, and is described in detail in Ref.~\cite{Khachatryan:2015iwa}.
The effects of pileup on the isolation variables are accounted for~\cite{Cacciari:2007fd}.

A conversion-safe electron veto algorithm~\cite{Khachatryan:2015iwa} is used to reject
electrons. It discards events containing a
track with an energy deposit in the innermost layer of the pixel
detector which is not connected to a reconstructed conversion vertex
from a photon cluster in the ECAL.
Events are required to have exactly one isolated photon candidate with
$\pt>25\GeV$ and $\abs{\eta}<1.44$.

The missing transverse momentum \ptmiss is defined
as the negative vector sum of the momenta of all reconstructed PF objects in the transverse plane.
The magnitude of  \ptmiss is required to
be larger than 30\GeV.

In order to select well-isolated objects, the photon is required to be
separated from the \PQb\ jet, the light-flavor jet, and any muon
candidates by $\DR(X,\gamma)>0.5$,
where $X$ stands for \Pgm, \PQb jet, or light-flavor jet.  The angular separation between the muon candidate and jets is required to
satisfy $\DR(\Pgm,\mathrm{jets})>0.3$.
The signal region is defined as the region where all the requirements
described above are satisfied.

The top quark is reconstructed using the muon, the \PQb\ jet candidate, and
\ptmiss where the
transverse momentum of the neutrino, which escapes detection, is assumed to be
equal to \ptmiss.
The longitudinal component of the neutrino momentum is
obtained by constraining the invariant mass of the muon and the neutrino to the \PW\ boson mass, resulting
in a quadratic equation.
When the two solutions are real, the one with smaller absolute value is taken, while
when the solutions are complex the real part is taken~\cite{Chatrchyan:2011vp}. The top quark
momentum is reconstructed by combining the momenta of the \PQb\ jet and of
the reconstructed \PW\ boson.

The backgrounds from
\wajets, \zajets, diboson+$\gamma$, $s$- and $\PQt\PW$-channel single top quark
with a photon production are estimated using simulation.
The contribution of the misidentified photon background
is estimated from the measurement of  the $\pt$-dependent probability for a jet to be
reconstructed as a photon using the method described in
Refs.~\cite{Chatrchyan:2011rr,Khachatryan:2016vif}.
The misidentification probability is
calculated  based on the near independence
of the shower shape and photon isolation variables, which permits the
factorization of the misidentification
probability and hence its estimation outside the signal region.

Several sources of uncertainties are considered in this estimation.
The first comes from the
definition of the sideband region and is due to possible correlations between the photon shower shape and
photon isolation, that are assumed to be independent. The second
source is due to the small contribution of genuine photons in the
sideband region, which is estimated from simulation.
The last source of uncertainty arises from the limited number of events available in the various photon
\pt bins considered in the estimation of the
background from misidentified jets.

The uncertainties in the lepton trigger,
reconstruction, and identification efficiencies are evaluated using Drell--Yan
events~\cite{Chatrchyan:2014mua, Sirunyan:2018fpa,Khachatryan:2015hwa}.
The uncertainties from the photon selection efficiency and photon energy scale are obtained using
$\cPZ \rightarrow \Pe^{+}+\Pe^{-}$ events.
The photon energy scale uncertainty is found to be
$1\%$~\cite{Chatrchyan:2013fya, Sirunyan:2016wqt}. The uncertainty from
pileup modeling is evaluated by varying the total \Pp\Pp\ inelastic cross section by
4.6\%~\cite{Aaboud:2016mmw} and reweighting the simulation accordingly. The uncertainties in the jet energy
resolution and jet energy scale
(JES)~\cite{CMS:2017wyc,Khachatryan:2016kdb}, and \ptmiss are also considered. The  \PQb\ jet identification and misidentification corrections are applied as a function of
the \pt and $\eta$ of the jet~\cite{Sirunyan:2017ezt}. The systematic uncertainty in the
integrated luminosity is estimated to be 2.5\%~\cite{CMS:2016eto}.
The uncertainties in the \wajets, \zajets, $\mathrm{VV}\gamma$,
and single top quark contributions are based on the CMS measurements
from Refs.~\cite{Khachatryan:2016vif},~\cite{zgammacs},~\cite{diboson}, and~\cite{stcs}, respectively.
Several sources of uncertainties in the
signal modeling are taken into account. The uncertainty from the
variations of  renormalization and factorization scales is estimated
by comparing the simulated samples with scales halved and doubled with respect to
the initial value. The uncertainty from the PDF choice is evaluated using the replicas of the
 NNPDF3.0 PDF set. The uncertainty from parton shower modeling and hadronization is
obtained by comparing the results derived from \PYTHIA v8.212 and
\HERWIGpp v2.3~\cite{Bahr:2008pv}.

After the selection, the total
number of observed events is 2535 of which
$2401\pm 178$ events are expected from the SM background in
absence of $\PQt\gamma$j signal, where the
uncertainty includes both the statistical and systematic components.
The expected number of signal events from the SM is $154\pm24$.
The dominant background is \ttbara, and amounts to 55\% of the total background yield.
Event yields in data and simulation for all background processes are
presented in Table~\ref{yield} where the \ttbara yield and its uncertainty are
predicted from simulation.

\begin{table}
\centering
\caption{Event yields after the event selection in data and for each SM contribution. The
expected yields are presented with their statistical and systematic uncertainties
combined. }
\label{yield}
\begin{tabular}{llll}
     Process                                            &      Event  yield \\ \hline
  \ttbara                                 &         $1401 \pm 131 $              \\
   \wajets                            &              $329  \pm 78 $             \\
   \zajets                             &          $232  \pm 55 $             \\
   Misidentified photon                           &        $374 \pm 74$\\
   $\PQt\gamma$ ($s$- and $\PQt\PW$-channel)       &        $57  \pm 8  $               \\
   VV$\gamma$                                     &        $8  \pm 3 $               \\
   Total background                               &  $2401  \pm 178  $ \\
  Expected signal                                  & $154 \pm 24 $\\
 Total SM prediction                              & $2555  \pm 180 $ \\
    Data                                                  &   $2535$  \\
 \end{tabular}
\end{table}

\begin{figure*}[htb]
\centering
\includegraphics[height=0.36\textwidth]{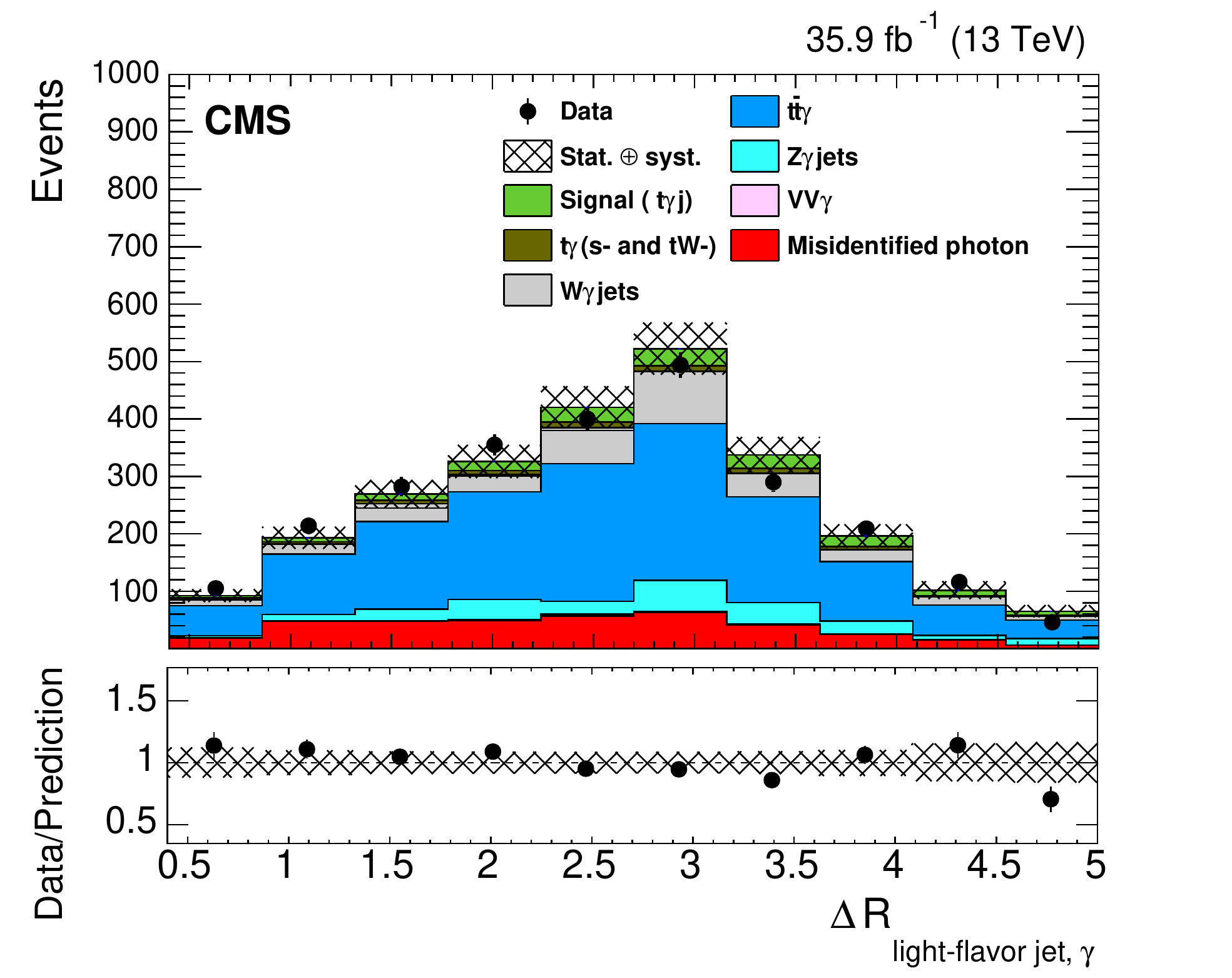}
\includegraphics[height=0.36\textwidth]{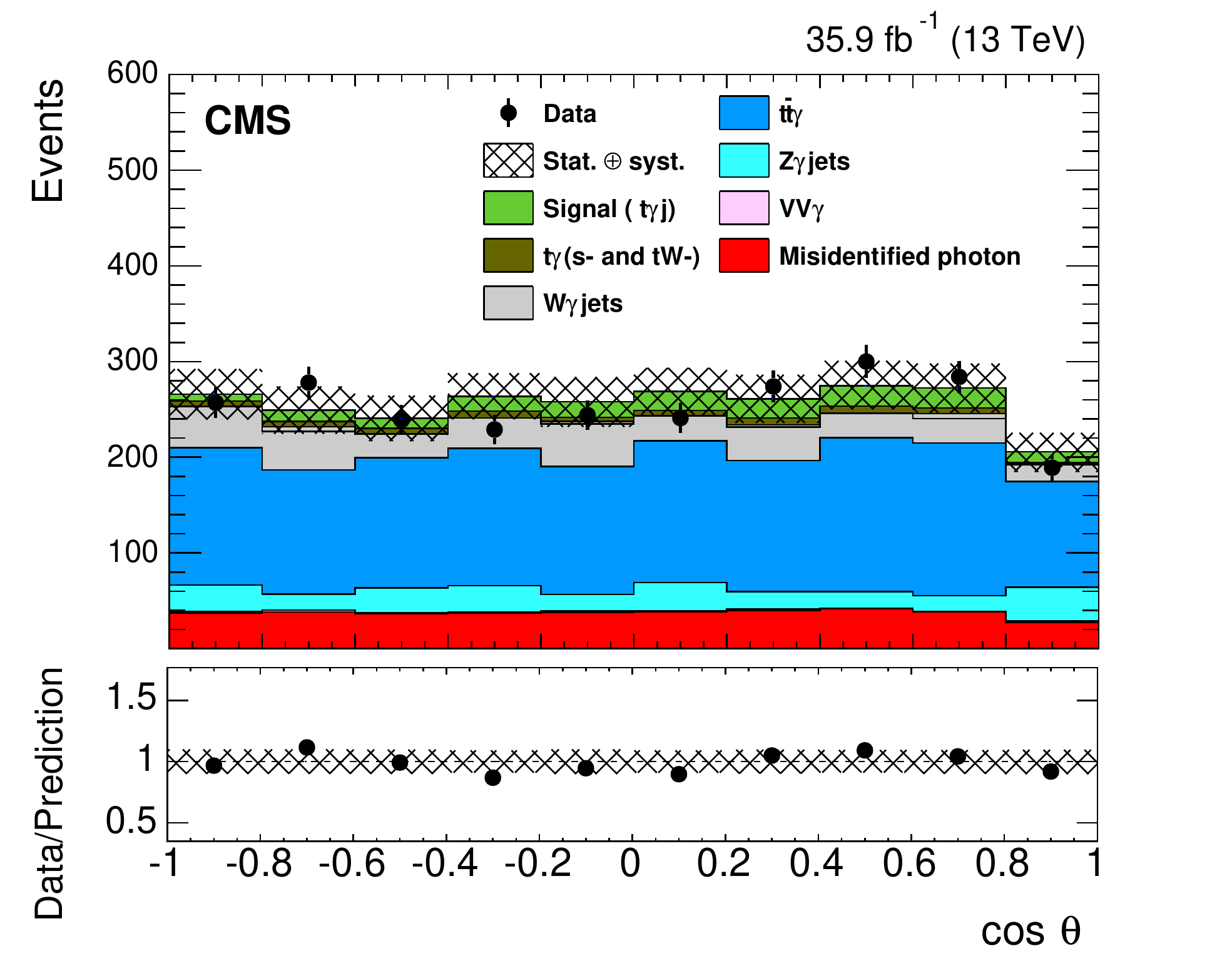}\\
\includegraphics[height=0.36\textwidth]{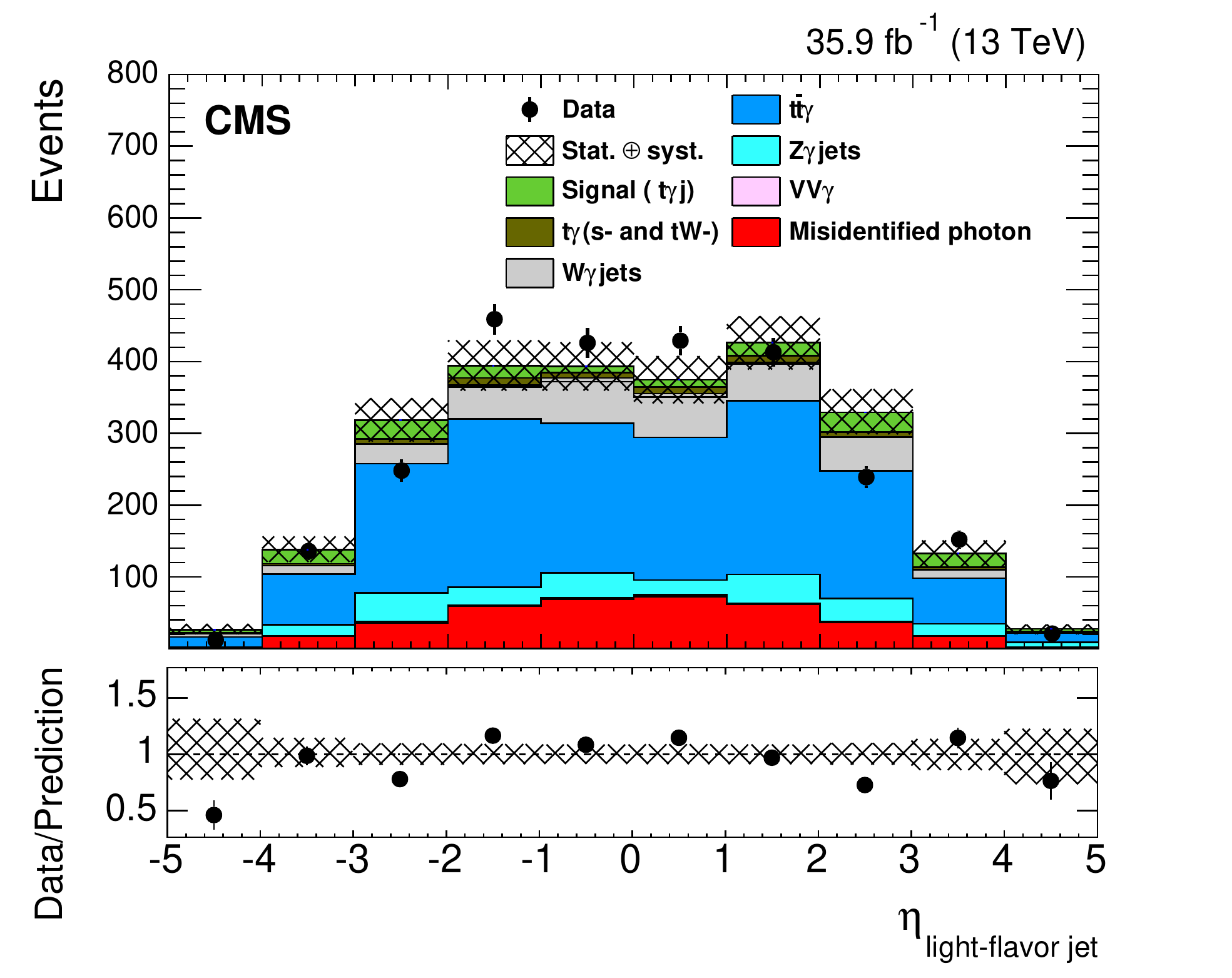}
\includegraphics[height=0.36\textwidth]{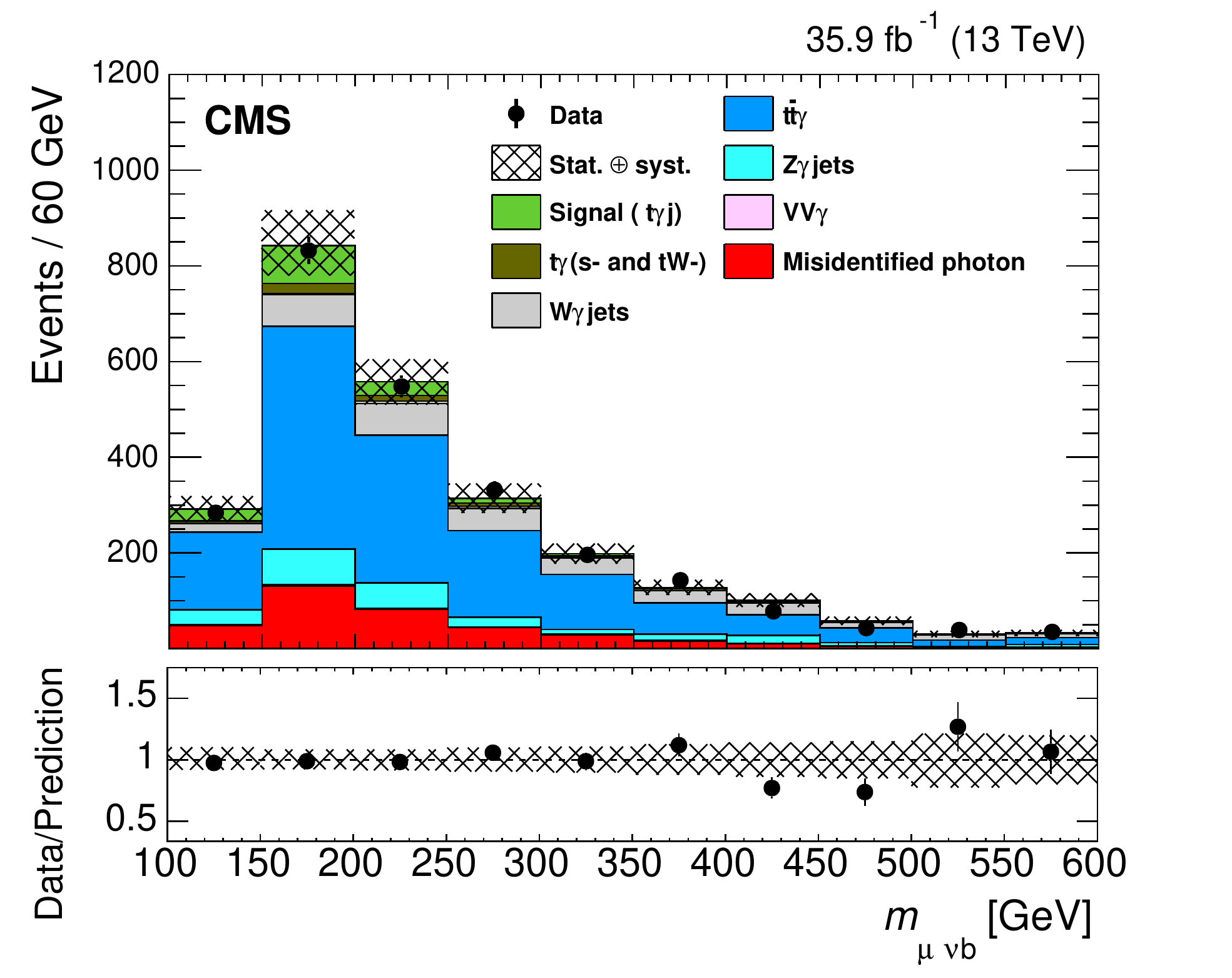}\\
\caption{Distributions of some of the input variables to the BDT: $\DR(\text{light jet},\gamma)$
  (upper left),  $\cos\theta$ (upper right), $\eta$ of the
  light-flavor jet (lower left), and $m_{\Pgm\nu \PQb}$ (lower right)
after the final event selection in data (points), and the SM prediction (filled histograms).
The hatched band shows the statistical and systematic
uncertainties in the estimated signal and background yields, and the
vertical bars on the points represent the statistical uncertainties of
the data. The ratios of the data to the SM predictions are shown in the bottom panels.
}
\label{photon}
\end{figure*}

The SM prediction for the signal is much smaller than the background, so
it is important to optimize the separation between signal and
background events. A multivariate
classification approach, based on a boosted decision tree (BDT)~\cite{Yang:2005nz,mva}, is used.
The BDT input variables in
decreasing order of importance are: (i)
$\eta$ of the light-flavor jet, (ii) $\cos\theta$, which is the cosine of the angle between the muon
candidate and the light-flavor jet in the top quark rest frame,
(iii)  $\eta$ of the muon candidate,  (iv) $\DR$(light-flavor jet,
$\gamma$), (v) reconstructed top quark mass $m_{\Pgm\nu \PQb}$, (vi)
jet multiplicity,
(vii) transverse mass of the reconstructed \PW\ boson, and (viii) muon
charge.
The distributions of some of the most important BDT input
variables for data and SM predictions
after the event selection are presented in
Fig.~\ref{photon}, where the hatched band indicates the overall
uncertainty on the SM predictions. It includes a sizable statistical
component besides the systematic uncertainties, which are all propagated
to the final results.

The BDT is trained to distinguish signal from background and provides
a single discriminant value for every event.
The BDT output distribution in data is parametrized as:
\begin{eqnarray}\label{param}
F(x) = C_{\mathrm{sig}}S_{\mathrm{sig}}(x) +C_{\ttbar\gamma}S_{\ttbar\gamma}(x)+
 C_{\PW\gamma\mathrm{j}} S_{\PW\gamma\mathrm{j}}(x)\nonumber \\ +C_{\cPZ\gamma\mathrm{j}}S_{\cPZ\gamma\mathrm{j}}(x)+
C_{\mathrm{misid}}S_{\mathrm{misid}}(x)+C_{\mathrm{ B}}S_{\mathrm{B}}(x),
\end{eqnarray}
where $x$ is the BDT output, $S_{\mathrm{sig}}(x)$, $S_{\PW\gamma\mathrm{j}}(x)$, $S_{\cPZ\gamma\mathrm{j}}(x)$,
$S_{\mathrm{misid}}(x)$, and $S_{\mathrm{B}}(x)$
are the normalized distributions (templates) for the signal,
\ttbara, \wajets, \zajets, the misidentified
photon background, and the sum of all other backgrounds which includes  single top quark production and $\text{VV}\gamma$,
respectively. The quantities $C_{\mathrm{sig}}$, $C_{\ttbar\gamma}$, $C_{\PW\gamma\mathrm{j}}$, $C_{\cPZ\gamma\mathrm{j}}$, $C_{\mathrm{misid}}$, and $C_{\mathrm{B}}$
are the corresponding normalizations  from which
$C_{\mathrm{sig}}$ and $C_{\ttbar\gamma}$ are left free in the fit, while the other terms are constrained by their associated uncertainties.
The different templates, $S(x)$, are taken from simulation except for
\ttbara and misidentified photons. The distribution $S_{\mathrm{misid}}(x)$ is obtained with the method described
above, calculating the yield as a function of BDT output. The template
$S_{\ttbar\gamma}(x)$ is
estimated from data using a control region defined by requiring exactly two $\PQb$-tagged jets,
while keeping all other selection criteria the same as for the signal region.
The requirement of two $\PQb$-tagged jets ensures a high contribution from
\ttbara,
while suppressing the contributions from all other processes.
The uncertainty in the difference of shape between events with two $\PQb$-tagged and one $\PQb$-tagged jet is
calculated using simulation and accounted for as a systematic uncertainty.
In addition, $S_{\PW\gamma\mathrm{j}}(x)$ is validated in data, in a control region enriched
in \wajets events, via the substitution of the signal \PQb\ jet requirement by a \PQb\ jet veto.

\begin{figure}[htb]
\centering
\includegraphics[width=0.49\textwidth]{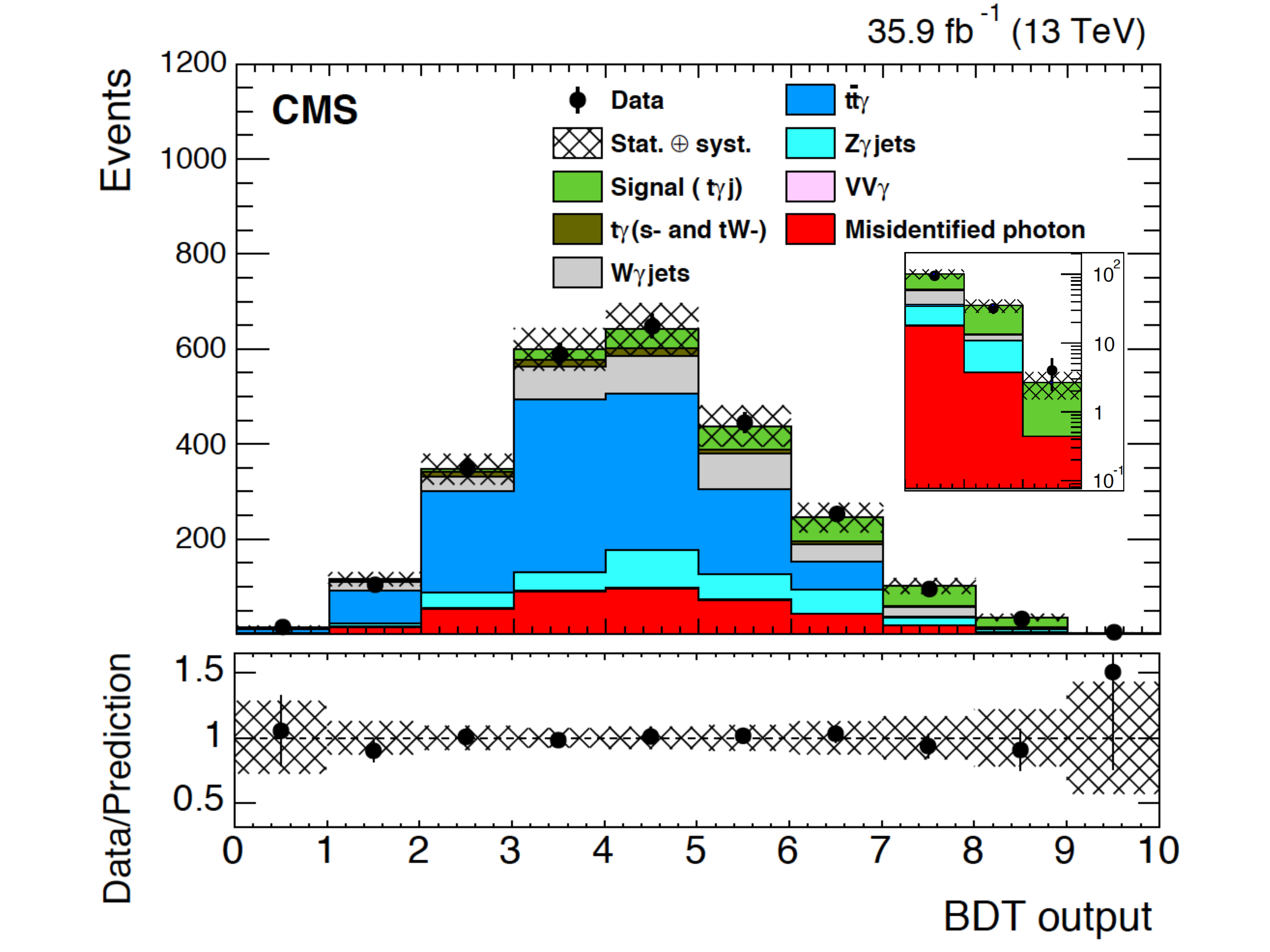}
\caption{ The BDT output  distribution for data and SM
  predictions after performing the fit. The inset
presents a closeup of the last three bins plotted on log scale. The hatched band shows the statistical and systematic
uncertainties in the estimated signal and background yields, and the
vertical bars on the points represent the statistical uncertainties of
the data. The ratio of the data to the SM
prediction is shown in the bottom panel.
}
\label{nnoutput}
\end{figure}

In order to extract the signal cross section and \ttbara background normalization, a simultaneous binned likelihood fit
is performed on the BDT distribution in the signal region and
the \ttbara control region. Including events from the \ttbara
control region in the fit is useful to constrain the \ttbara
background normalization. Each source of systematic uncertainty is included
as a nuisance parameter in the likelihood function. The normalizations
of backgrounds except for \ttbara
 are left free to vary within the systematic uncertainties.
A profile likelihood ratio test statistic is constructed by generating pseudo-data  for the background-only and for the signal-plus-background
hypotheses. The BDT output distribution for data and SM
prediction after the fit is shown in Fig.~\ref{nnoutput}.

All of the systematic uncertainties affect both the normalization of
backgrounds and shape of the BDT discriminant,
except those associated with the integrated luminosity, photon energy scale, \ptmiss, and
background rates that only affect the normalization. The shape uncertainties have Gaussian constraints,
while rate uncertainties have log-normal forms.
The main systematic uncertainties in the signal cross section arise
from the  JES, signal
modeling,  normalization of \zajets, and \PQb\ tagging and mistagging rates, and
amount to 12, 9, 8, and 7\%,
respectively. The impact of the uncertainty from each source is calculated by performing the fit with
all other nuisance parameters fixed to their fitted values. The number
of signal and \ttbara, events after the fit are $220\pm63$ and $1221\pm121$, which both
agree with the expected yields within the uncertainties.

An excess of events above the expected background is observed at a
$p$-value~\cite{Lyons:2008zz} of $4.27\times 10^{-6}$, which corresponds to
a significance of 4.4 standard deviations.
The median expected significance is 3.0, and the 68 and 95\%
confidence level ranges for the expected
significance are $[1.5,4.0]$ and $[0,8.7]$, respectively.
A fiducial product of the cross section and branching fraction of $\sigma(\Pp\Pp\to\PQt\gamma\mathrm{j})\mathcal{B}(\PQt\to\Pgm\nu\PQb) = 115\pm 17\stat\pm 30\syst\unit{fb}$ is measured in the phase space defined by the photon transverse momentum  $p_{{\mathrm {T}},\gamma} > 25$\GeV,
$\abs{\eta_{\gamma}}<1.44$, and $\DR(X,\gamma) > 0.5$, where $X$ stands for \Pgm, \PQb\ jet, light-flavor jet.
The expected SM product of the cross section and branching fraction within this fiducial phase space is $81\pm4\unit{fb}$,  in agreement
with the measurement. This is the first experimental
evidence for single top quark production in association with a photon.

\begin{acknowledgments}
We congratulate our colleagues in the CERN accelerator departments for
the excellent performance of the LHC and thank the technical and
administrative staffs at CERN and at other CMS institutes for their
contributions to the success of the CMS effort. In addition, we
gratefully acknowledge the computing centres and personnel of the
Worldwide LHC Computing Grid for delivering so effectively the
computing infrastructure essential to our analyses. Finally, we
acknowledge the enduring support for the construction and operation of
the LHC and the CMS detector provided by the following funding
agencies: BMWFW and FWF (Austria); FNRS and FWO (Belgium); CNPq,
CAPES, FAPERJ, and FAPESP (Brazil); MES (Bulgaria); CERN; CAS, MoST,
and NSFC (China); COLCIENCIAS (Colombia); MSES and CSF (Croatia); RPF
(Cyprus); SENESCYT (Ecuador); MoER, ERC IUT, and ERDF (Estonia);
Academy of Finland, MEC, and HIP (Finland); CEA and CNRS/IN2P3
(France); BMBF, DFG, and HGF (Germany); GSRT (Greece); OTKA and NIH
(Hungary); DAE and DST (India); IPM (Iran); SFI (Ireland); INFN
(Italy); MSIP and NRF (Republic of Korea); LAS (Lithuania); MOE and UM
(Malaysia); BUAP, CINVESTAV, CONACYT, LNS, SEP, and UASLP-FAI
(Mexico); MBIE (New Zealand); PAEC (Pakistan); MSHE and NSC (Poland);
FCT (Portugal); JINR (Dubna); MON, RosAtom, RAS, RFBR and RAEP
(Russia); MESTD (Serbia); SEIDI, CPAN, PCTI and FEDER (Spain); Swiss
Funding Agencies (Switzerland); MST (Taipei); ThEPCenter, IPST, STAR,
and NSTDA (Thailand); TUBITAK and TAEK (Turkey); NASU and SFFR
(Ukraine); STFC (United Kingdom); DOE and NSF (USA).
Individuals have received support from INSF (Iran).
\end{acknowledgments}

\bibliography{auto_generated}
\cleardoublepage \appendix\section{The CMS Collaboration \label{app:collab}}\begin{sloppypar}\hyphenpenalty=5000\widowpenalty=500\clubpenalty=5000\input{TOP-17-016-authorlist.tex}\end{sloppypar}
\end{document}

%% file: TOP-17-016-authorlist.tex
\vskip\cmsinstskip
\textbf{Yerevan Physics Institute, Yerevan, Armenia}\\*[0pt]
A.M.~Sirunyan, A.~Tumasyan
\vskip\cmsinstskip
\textbf{Institut f\"{u}r Hochenergiephysik, Wien, Austria}\\*[0pt]
W.~Adam, F.~Ambrogi, E.~Asilar, T.~Bergauer, J.~Brandstetter, M.~Dragicevic, J.~Er\"{o}, A.~Escalante~Del~Valle, M.~Flechl, R.~Fr\"{u}hwirth\cmsAuthorMark{1}, V.M.~Ghete, J.~Hrubec, M.~Jeitler\cmsAuthorMark{1}, N.~Krammer, I.~Kr\"{a}tschmer, D.~Liko, T.~Madlener, I.~Mikulec, N.~Rad, H.~Rohringer, J.~Schieck\cmsAuthorMark{1}, R.~Sch\"{o}fbeck, M.~Spanring, D.~Spitzbart, A.~Taurok, W.~Waltenberger, J.~Wittmann, C.-E.~Wulz\cmsAuthorMark{1}, M.~Zarucki
% ...
\vskip\cmsinstskip
% ...
1:  Also at Vienna University of Technology, Vienna, Austria\\
% ...


%% file: TOP-17-016_temp.bbl
\providecommand{\href}[2]{#2}\begingroup\raggedright\begin{thebibliography}{10}%
\makeatletter
\providecommand{\hrefCMSnoop }[0]{\@secondoftwo}%
\makeatother
\providecommand{\doi}{\texttt{doi:}\begingroup \urlstyle{tt}\Url}

\bibitem{Fael:2013ira}
\hrefCMSnoop {}{M.~Fael and T.~Gehrmann, ``{Probing top quark electromagnetic
  dipole moments in single-top-plus-photon production}'',} \textit{ Phys. Rev.
  D} \textbf{ 88} (2013) 033003,
  \href{http://dx.doi.org/10.1103/PhysRevD.88.033003}{\doi{10.1103/PhysRevD.88.033003}},
\href{http://www.arXiv.org/abs/1307.1349}{\texttt{arXiv:1307.1349}}.

\bibitem{Etesami:2016rwu}
\hrefCMSnoop {}{S.~M. Etesami, S.~Khatibi, and M.~Mohammadi~Najafabadi,
  ``{Measuring anomalous WW $\gamma $ and t$\bar{\text {t}}\gamma $ couplings
  using top+$\gamma $ production at the LHC}'',} \textit{ Eur. Phys. J. C}
  \textbf{ 76} (2016) 533,
  \href{http://dx.doi.org/10.1140/epjc/s10052-016-4376-2}{\doi{10.1140/epjc/s10052-016-4376-2}},
\href{http://www.arXiv.org/abs/1606.02178}{\texttt{arXiv:1606.02178}}.

\bibitem{Aaltonen:2011sp}
\hrefCMSnoop {}{{CDF} Collaboration, ``{Evidence for $t\bar{t}\gamma$
  production and measurement of $\sigma_{t\bar{t}\gamma} /
  \sigma_{t\bar{t}}$}'',} \textit{ Phys. Rev. D} \textbf{ 84} (2011) 031104,
  \href{http://dx.doi.org/10.1103/PhysRevD.84.031104}{\doi{10.1103/PhysRevD.84.031104}},
\href{http://www.arXiv.org/abs/1106.3970}{\texttt{arXiv:1106.3970}}.

\bibitem{Aaboud:2017era}
\hrefCMSnoop {}{{ATLAS Collaboration}, ``{Measurement of the $t\bar{t}\gamma$
  production cross section in proton-proton collisions at $\sqrt{s} = 8$ TeV
  with the ATLAS detector}'',} \textit{ JHEP} \textbf{ 11} (2017) 086,
  \href{http://dx.doi.org/10.1007/JHEP11(2017)086}{\doi{10.1007/JHEP11(2017)086}},
\href{http://www.arXiv.org/abs/1706.03046}{\texttt{arXiv:1706.03046}}.

\bibitem{Sirunyan:2017iyh}
\hrefCMSnoop {}{{CMS Collaboration}, ``{Measurement of the semileptonic $
  \mathrm{t}\overline{\mathrm{t}}+\gamma$ production cross section in pp
  collisions at $ \sqrt{s}=8 $ TeV}'',} \textit{ JHEP} \textbf{ 10} (2017) 006,
  \href{http://dx.doi.org/10.1007/JHEP10(2017)006}{\doi{10.1007/JHEP10(2017)006}},
\href{http://www.arXiv.org/abs/1706.08128}{\texttt{arXiv:1706.08128}}.

\bibitem{trigger}
\hrefCMSnoop {}{{CMS Collaboration}, ``{The CMS trigger system}'',} \textit{
  JINST} \textbf{ 12} (2017) P01020,
  \href{http://dx.doi.org/10.1088/1748-0221/12/01/P01020}{\doi{10.1088/1748-0221/12/01/P01020}},
\href{http://www.arXiv.org/abs/1609.02366}{\texttt{arXiv:1609.02366}}.

\bibitem{Chatrchyan:2008zzk}
\hrefCMSnoop {}{{CMS Collaboration}, ``The {CMS} experiment at the {CERN}
  {LHC}'',} \textit{ JINST} \textbf{ 3} (2008) S08004,
\href{http://dx.doi.org/10.1088/1748-0221/3/08/S08004}{\doi{10.1088/1748-0221/3/08/S08004}}.

\bibitem{Alwall:2014hca}
J.~Alwall\hrefCMSnoop {}{ {et~al.}, ``{The automated computation of tree-level
  and next-to-leading order differential cross sections, and their matching to
  parton shower simulations}'',} \textit{ JHEP} \textbf{ 07} (2014) 079,
  \href{http://dx.doi.org/10.1007/JHEP07(2014)079}{\doi{10.1007/JHEP07(2014)079}},
\href{http://www.arXiv.org/abs/1405.0301}{\texttt{arXiv:1405.0301}}.

\bibitem{Ball:2014uwa}
\hrefCMSnoop {}{{NNPDF} Collaboration, ``{Parton distributions for the LHC Run
  II}'',} \textit{ JHEP} \textbf{ 04} (2015) 040,
  \href{http://dx.doi.org/10.1007/JHEP04(2015)040}{\doi{10.1007/JHEP04(2015)040}},
\href{http://www.arXiv.org/abs/1410.8849}{\texttt{arXiv:1410.8849}}.

\bibitem{Kalogeropoulos:2018cke}
\hrefCMSnoop {}{A.~Kalogeropoulos and J.~Alwall, ``{The SysCalc code: A tool to
  derive theoretical systematic uncertainties}'',} (2018).
\href{http://www.arXiv.org/abs/1801.08401}{\texttt{arXiv:1801.08401}}.

\bibitem{Nason:2004rx}
\hrefCMSnoop {}{P.~Nason, ``{A new method for combining NLO QCD with shower
  Monte Carlo algorithms}'',} \textit{ JHEP} \textbf{ 11} (2004) 040,
  \href{http://dx.doi.org/10.1088/1126-6708/2004/11/040}{\doi{10.1088/1126-6708/2004/11/040}},
\href{http://www.arXiv.org/abs/hep-ph/0409146}{\texttt{arXiv:hep-ph/0409146}}.

\bibitem{Frixione:2007vw}
\hrefCMSnoop {}{S.~Frixione, P.~Nason, and C.~Oleari, ``{Matching NLO QCD
  computations with parton shower simulations: the POWHEG method}'',} \textit{
  JHEP} \textbf{ 11} (2007) 070,
  \href{http://dx.doi.org/10.1088/1126-6708/2007/11/070}{\doi{10.1088/1126-6708/2007/11/070}},
\href{http://www.arXiv.org/abs/0709.2092}{\texttt{arXiv:0709.2092}}.

\bibitem{Alioli:2010xd}
\hrefCMSnoop {}{S.~Alioli, P.~Nason, C.~Oleari, and E.~Re, ``{A general
  framework for implementing NLO calculations in shower Monte Carlo programs:
  the POWHEG BOX}'',} \textit{ JHEP} \textbf{ 06} (2010) 043,
  \href{http://dx.doi.org/10.1007/JHEP06(2010)043}{\doi{10.1007/JHEP06(2010)043}},
\href{http://www.arXiv.org/abs/1002.2581}{\texttt{arXiv:1002.2581}}.

\bibitem{Campbell:2014kua}
\hrefCMSnoop {}{J.~M. Campbell, R.~K. Ellis, P.~Nason, and E.~Re, ``{Top-pair
  production and decay at NLO matched with parton showers}'',} \textit{ JHEP}
  \textbf{ 04} (2015) 114,
  \href{http://dx.doi.org/10.1007/JHEP04(2015)114}{\doi{10.1007/JHEP04(2015)114}},
\href{http://www.arXiv.org/abs/1412.1828}{\texttt{arXiv:1412.1828}}.

\bibitem{Sjostrand:2006za}
\hrefCMSnoop {}{T.~Sj{\"o}strand, S.~Mrenna, and P.~Z. Skands, ``{PYTHIA 6.4
  Physics and Manual}'',} \textit{ JHEP} \textbf{ 05} (2006) 026,
  \href{http://dx.doi.org/10.1088/1126-6708/2006/05/026}{\doi{10.1088/1126-6708/2006/05/026}},
\href{http://www.arXiv.org/abs/hep-ph/0603175}{\texttt{arXiv:hep-ph/0603175}}.

\bibitem{Sjostrand:2007gs}
\hrefCMSnoop {}{T.~Sj{\"o}strand, S.~Mrenna, and P.~Z. Skands, ``{A Brief
  Introduction to PYTHIA 8.1}'',} \textit{ Comput. Phys. Commun.} \textbf{ 178}
  (2008) 852,
  \href{http://dx.doi.org/10.1016/j.cpc.2008.01.036}{\doi{10.1016/j.cpc.2008.01.036}},
\href{http://www.arXiv.org/abs/0710.3820}{\texttt{arXiv:0710.3820}}.

\bibitem{Khachatryan:2015pea}
\hrefCMSnoop {}{{CMS Collaboration}, ``{Event generator tunes obtained from
  underlying event and multiparton scattering measurements}'',} \textit{ Eur.
  Phys. J. C} \textbf{ 76} (2016) 155,
  \href{http://dx.doi.org/10.1140/epjc/s10052-016-3988-x}{\doi{10.1140/epjc/s10052-016-3988-x}},
\href{http://www.arXiv.org/abs/1512.00815}{\texttt{arXiv:1512.00815}}.

\bibitem{Alwall:2007fs}
\hrefCMSnoop {}{J.~Alwall {et~al.}, ``{Comparative study of various algorithms
  for the merging of parton showers and matrix elements in hadronic
  collisions}'',} \textit{ Eur. Phys. J.} \textbf{ C53} (2008) 473,
  \href{http://dx.doi.org/10.1140/epjc/s10052-007-0490-5}{\doi{10.1140/epjc/s10052-007-0490-5}},
\href{http://www.arXiv.org/abs/0706.2569}{\texttt{arXiv:0706.2569}}.

\bibitem{Frederix:2012ps}
\hrefCMSnoop {}{R.~Frederix and S.~Frixione, ``{Merging meets matching in
  MC@NLO}'',} \textit{ JHEP} \textbf{ 12} (2012) 061,
  \href{http://dx.doi.org/10.1007/JHEP12(2012)061}{\doi{10.1007/JHEP12(2012)061}},
\href{http://www.arXiv.org/abs/1209.6215}{\texttt{arXiv:1209.6215}}.

\bibitem{geant}
\hrefCMSnoop {}{{GEANT4} Collaboration, ``{GEANT4}---a simulation toolkit'',}
  \textit{ Nucl. Instrum. Meth. A} \textbf{ 506} (2003) 250,
\href{http://dx.doi.org/10.1016/S0168-9002(03)01368-8}{\doi{10.1016/S0168-9002(03)01368-8}}.

\bibitem{Sirunyan:2017ulk}
\hrefCMSnoop {}{{CMS Collaboration}, ``{Particle-flow reconstruction and global
  event description with the CMS detector}'',} \textit{ JINST} \textbf{ 12}
  (2017) P10003,
  \href{http://dx.doi.org/10.1088/1748-0221/12/10/P10003}{\doi{10.1088/1748-0221/12/10/P10003}},
\href{http://www.arXiv.org/abs/1706.04965}{\texttt{arXiv:1706.04965}}.

\bibitem{Chatrchyan:2014fea}
\hrefCMSnoop {}{{CMS Collaboration}, ``{Description and performance of track
  and primary-vertex reconstruction with the CMS tracker}'',} \textit{ JINST}
  \textbf{ 9} (2014) P10009,
  \href{http://dx.doi.org/10.1088/1748-0221/9/10/P10009}{\doi{10.1088/1748-0221/9/10/P10009}},
\href{http://www.arXiv.org/abs/1405.6569}{\texttt{arXiv:1405.6569}}.

\bibitem{Cacciari:2008gp}
\hrefCMSnoop {}{M.~Cacciari, G.~P. Salam, and G.~Soyez, ``The anti-\kt jet
  clustering algorithm'',} \textit{ JHEP} \textbf{ 04} (2008) 063,
  \href{http://dx.doi.org/10.1088/1126-6708/2008/04/063}{\doi{10.1088/1126-6708/2008/04/063}},
  \href{http://www.arXiv.org/abs/0802.1189}{\texttt{arXiv:0802.1189}}.

\bibitem{Cacciari:2011ma}
\hrefCMSnoop {}{M.~Cacciari, G.~P. Salam, and G.~Soyez, ``{FastJet user
  manual}'',} \textit{ Eur. Phys. J. C} \textbf{ 72} (2012) 1896,
  \href{http://dx.doi.org/10.1140/epjc/s10052-012-1896-2}{\doi{10.1140/epjc/s10052-012-1896-2}},
\href{http://www.arXiv.org/abs/1111.6097}{\texttt{arXiv:1111.6097}}.

\bibitem{Sirunyan:2017ezt}
\hrefCMSnoop {}{{CMS Collaboration}, ``{Identification of heavy-flavour jets
  with the CMS detector in pp collisions at 13 TeV}'',} \textit{ JINST}
  \textbf{ 13} (2018) P05011,
  \href{http://dx.doi.org/10.1088/1748-0221/13/05/P05011}{\doi{10.1088/1748-0221/13/05/P05011}},
\href{http://www.arXiv.org/abs/1712.07158}{\texttt{arXiv:1712.07158}}.

\bibitem{Sirunyan:2018fpa}
\hrefCMSnoop {}{{CMS Collaboration}, ``{Performance of the CMS muon detector
  and muon reconstruction with proton-proton collisions at $\sqrt{s}=$ 13
  TeV}'',} \textit{ JINST} \textbf{ 13} (2018) P06015,
  \href{http://dx.doi.org/10.1088/1748-0221/13/06/P06015}{\doi{10.1088/1748-0221/13/06/P06015}},
\href{http://www.arXiv.org/abs/1804.04528}{\texttt{arXiv:1804.04528}}.

\bibitem{Khachatryan:2015iwa}
\hrefCMSnoop {}{{CMS Collaboration}, ``{Performance of photon reconstruction
  and identification with the CMS detector in proton-proton collisions at
  $\sqrt{s}=8$ TeV}'',} \textit{ JINST} \textbf{ 10} (2015) P08010,
  \href{http://dx.doi.org/10.1088/1748-0221/10/08/P08010}{\doi{10.1088/1748-0221/10/08/P08010}},
\href{http://www.arXiv.org/abs/1502.02702}{\texttt{arXiv:1502.02702}}.

\bibitem{Cacciari:2007fd}
\hrefCMSnoop {}{M.~Cacciari and G.~P. Salam, ``{Pileup subtraction using jet
  areas}'',} \textit{ Phys. Lett. B} \textbf{ 659} (2008) 119,
  \href{http://dx.doi.org/10.1016/j.physletb.2007.09.077}{\doi{10.1016/j.physletb.2007.09.077}},
\href{http://www.arXiv.org/abs/0707.1378}{\texttt{arXiv:0707.1378}}.

\bibitem{Chatrchyan:2011vp}
\hrefCMSnoop {}{{CMS Collaboration}, ``{Measurement of the $t$-channel single
  top quark production cross section in pp collisions at $\sqrt{s}=7$ TeV}'',}
  \textit{ Phys. Rev. Lett.} \textbf{ 107} (2011) 091802,
  \href{http://dx.doi.org/10.1103/PhysRevLett.107.091802}{\doi{10.1103/PhysRevLett.107.091802}},
\href{http://www.arXiv.org/abs/1106.3052}{\texttt{arXiv:1106.3052}}.

\bibitem{Chatrchyan:2011rr}
\hrefCMSnoop {}{{CMS Collaboration}, ``{Measurement of $W\gamma$ and $Z\gamma$
  production in pp collisions at $\sqrt{s} =$ 7 TeV}'',} \textit{ Phys. Lett.
  B} \textbf{ 701} (2011) 535,
  \href{http://dx.doi.org/10.1016/j.physletb.2011.06.034}{\doi{10.1016/j.physletb.2011.06.034}},
\href{http://www.arXiv.org/abs/1105.2758}{\texttt{arXiv:1105.2758}}.

\bibitem{Khachatryan:2016vif}
\hrefCMSnoop {}{{CMS Collaboration}, ``{Measurement of electroweak-induced
  production of W$\gamma$ with two jets in pp collisions at $ \sqrt{s}=$ 8 TeV
  and constraints on anomalous quartic gauge couplings}'',} \textit{ JHEP}
  \textbf{ 06} (2017) 106,
  \href{http://dx.doi.org/10.1007/JHEP06(2017)106}{\doi{10.1007/JHEP06(2017)106}},
\href{http://www.arXiv.org/abs/1612.09256}{\texttt{arXiv:1612.09256}}.

\bibitem{Chatrchyan:2014mua}
\hrefCMSnoop {}{{CMS Collaboration}, ``{Measurement of inclusive W and Z boson
  production cross sections in pp collisions at $\sqrt{s}$ = 8 TeV}'',}
  \textit{ Phys. Rev. Lett.} \textbf{ 112} (2014) 191802,
  \href{http://dx.doi.org/10.1103/PhysRevLett.112.191802}{\doi{10.1103/PhysRevLett.112.191802}},
\href{http://www.arXiv.org/abs/1402.0923}{\texttt{arXiv:1402.0923}}.

\bibitem{Khachatryan:2015hwa}
\hrefCMSnoop {}{{CMS Collaboration}, ``{Performance of electron reconstruction
  and selection with the CMS detector in proton-proton collisions at $\sqrt{s}
  = 8$ TeV}'',} \textit{ JINST} \textbf{ 10} (2015) P06005,
  \href{http://dx.doi.org/10.1088/1748-0221/10/06/P06005}{\doi{10.1088/1748-0221/10/06/P06005}},
\href{http://www.arXiv.org/abs/1502.02701}{\texttt{arXiv:1502.02701}}.

\bibitem{Chatrchyan:2013fya}
\hrefCMSnoop {}{{CMS Collaboration}, ``{Measurement of the $W\gamma$ and
  $Z\gamma$ inclusive cross sections in pp collisions at $\sqrt{s}= $ 7 TeV and
  limits on anomalous triple gauge boson couplings}'',} \textit{ Phys. Rev. D}
  \textbf{ 89} (2014) 092005,
  \href{http://dx.doi.org/10.1103/PhysRevD.89.092005}{\doi{10.1103/PhysRevD.89.092005}},
\href{http://www.arXiv.org/abs/1308.6832}{\texttt{arXiv:1308.6832}}.

\bibitem{Sirunyan:2016wqt}
\hrefCMSnoop {}{{CMS Collaboration}, ``{Search for high-mass $\mathrm{ Z
  }\gamma$ resonances in proton-proton collisions at $\sqrt{s}=$ 8 and 13 TeV
  using jet substructure techniques}'',} \textit{ Phys. Lett.} \textbf{ B772}
  (2017) 363,
  \href{http://dx.doi.org/10.1016/j.physletb.2017.06.062}{\doi{10.1016/j.physletb.2017.06.062}},
\href{http://www.arXiv.org/abs/1612.09516}{\texttt{arXiv:1612.09516}}.

\bibitem{Aaboud:2016mmw}
\hrefCMSnoop {}{{ATLAS Collaboration}, ``{Measurement of the inelastic
  proton-proton cross section at $\sqrt{s} = 13$ TeV with the ATLAS detector at
  the LHC}'',} \textit{ Phys. Rev. Lett.} \textbf{ 117} (2016) 182002,
  \href{http://dx.doi.org/10.1103/PhysRevLett.117.182002}{\doi{10.1103/PhysRevLett.117.182002}},
\href{http://www.arXiv.org/abs/1606.02625}{\texttt{arXiv:1606.02625}}.

\bibitem{CMS:2017wyc}
\href {https://cds.cern.ch/record/2256875?ln=en}{{CMS Collaboration}, ``{Jet
  algorithms performance in 13 TeV data}'',} CMS Physics Analysis Summary
  CMS-PAS-JME-16-003, 2017.

\bibitem{Khachatryan:2016kdb}
\hrefCMSnoop {}{{CMS Collaboration}, ``{Jet energy scale and resolution in the
  CMS experiment in pp collisions at 8 TeV}'',} \textit{ JINST} \textbf{ 12}
  (2017) P02014,
  \href{http://dx.doi.org/10.1088/1748-0221/12/02/P02014}{\doi{10.1088/1748-0221/12/02/P02014}},
\href{http://www.arXiv.org/abs/1607.03663}{\texttt{arXiv:1607.03663}}.

\bibitem{CMS:2016eto}
\href {https://cds.cern.ch/record/2257069?ln=en}{{CMS Collaboration}, ``{CMS
  luminosity measurements for the 2016 data taking period}'',} CMS Physics
  Analysis Summary CMS-PAS-LUM-17-001, 2017.

\bibitem{zgammacs}
\hrefCMSnoop {}{{CMS Collaboration}, ``{Measurement of the cross section for
  electroweak production of Z$\gamma$ in association with two jets and
  constraints on anomalous quartic gauge couplings in proton-proton collisions
  at $\sqrt{s} = 8$ TeV}'',} \textit{ Phys. Lett. B} \textbf{ 770} (2017) 380,
  \href{http://dx.doi.org/10.1016/j.physletb.2017.04.071}{\doi{10.1016/j.physletb.2017.04.071}},
\href{http://www.arXiv.org/abs/1702.03025}{\texttt{arXiv:1702.03025}}.

\bibitem{diboson}
\hrefCMSnoop {}{{CMS Collaboration}, ``{Measurement of the ${{\mathrm{W} }^{+}
  }\mathrm{W}^{-} $ cross section in pp collisions at $\sqrt{s} =$ 8 TeV and
  limits on anomalous gauge couplings}'',} \textit{ Eur. Phys. J. C} \textbf{
  76} (2016) 401,
  \href{http://dx.doi.org/10.1140/epjc/s10052-016-4219-1}{\doi{10.1140/epjc/s10052-016-4219-1}},
\href{http://www.arXiv.org/abs/1507.03268}{\texttt{arXiv:1507.03268}}.

\bibitem{stcs}
\hrefCMSnoop {}{{CMS Collaboration}, ``{Cross section measurement of
  $t$-channel single top quark production in pp collisions at $\sqrt{s} =$ 13
  TeV}'',} \textit{ Phys. Lett. B} \textbf{ 772} (2017) 752,
  \href{http://dx.doi.org/10.1016/j.physletb.2017.07.047}{\doi{10.1016/j.physletb.2017.07.047}},
\href{http://www.arXiv.org/abs/1610.00678}{\texttt{arXiv:1610.00678}}.

\bibitem{Bahr:2008pv}
\hrefCMSnoop {}{M.~B{\"a}hr {et~al.}, ``{\HERWIGpp physics and manual}'',}
  \textit{ Eur. Phys. J. C} \textbf{ 58} (2008) 639,
  \href{http://dx.doi.org/10.1140/epjc/s10052-008-0798-9}{\doi{10.1140/epjc/s10052-008-0798-9}},
\href{http://www.arXiv.org/abs/0803.0883}{\texttt{arXiv:0803.0883}}.

\bibitem{Yang:2005nz}
\hrefCMSnoop {}{H.-J. Yang, B.~P. Roe, and J.~Zhu, ``{Studies of boosted
  decision trees for MiniBooNE particle identification}'',} \textit{ Nucl.
  Instrum. Meth. A} \textbf{ 555} (2005) 370,
  \href{http://dx.doi.org/10.1016/j.nima.2005.09.022}{\doi{10.1016/j.nima.2005.09.022}},
\href{http://www.arXiv.org/abs/physics/0508045}{\texttt{arXiv:physics/0508045}}.

\bibitem{mva}
\href {http://pos.sissa.it/archive/conferences/050/040/ACAT_040.pdf}{H.~Voss,
  A.~H{\"o}cker, J.~Stelzer, and F.~Tegenfeldt, ``{TMVA}, the toolkit for
  multivariate data analysis with {ROOT}'',} in \textit{ XIth International
  Workshop on Advanced Computing and Analysis Techniques in Physics Research
  (ACAT)}, p.~40.
\newblock 2007.
\newblock
\href{http://www.arXiv.org/abs/physics/0703039}{\texttt{arXiv:physics/0703039}}.
\newblock

\bibitem{Lyons:2008zz}
\hrefCMSnoop {}{L.~Lyons, H.~B. Prosper, and A.~De~Roeck, eds., ``{Statistical
  issues for LHC physics. Proceedings, Workshop, PHYSTAT-LHC, Geneva,
  Switzerland, June 27-29, 2007}''}.
\newblock (2008).
\newblock
\href{http://dx.doi.org/10.5170/CERN-2008-001}{\doi{10.5170/CERN-2008-001}}.

\end{thebibliography}\endgroup
